\documentclass[%
 aip,
 amsmath,amssymb,
 reprint,%
]{revtex4-1}

\usepackage{graphicx}
\usepackage{dcolumn}
\usepackage{bm}

\usepackage[utf8]{inputenc}
\usepackage[T1]{fontenc}
\usepackage{mathptmx}
\usepackage{etoolbox}
\usepackage{multirow}
\usepackage{booktabs} 
\usepackage{subfig} 

\makeatletter
\def\@email#1#2{%
 \endgroup
 \patchcmd{\titleblock@produce}
  {\frontmatter@RRAPformat}
  {\frontmatter@RRAPformat{\produce@RRAP{*#1\href{mailto:#2}{#2}}}\frontmatter@RRAPformat}
  {}{}
}%
\makeatother
\begin{document}

\preprint{AIP/123-QED}

\title{Physics-Informed Neural Networks for the Korteweg-de Vries Equation for Internal Solitary Wave Problem: Forward Simulation and Inverse Parameter Estimation}

\author{Ming Kang}%
\affiliation{School of Science, China University of Geosciences (Beijing), Beijing, China}%

\author{Hang Li}%
\affiliation{School of Integrated Circuits and Electronics, Beijing Institute of Technology, Beijing, China}%

\author{Qiwen Tan}
\affiliation{Science42 Technology, Beijing, China}%

\author{Zhan Wang}
\affiliation{College of Meteorology and Oceanography, National University of Defense Technology, Changsha, China}

\author{Ruipeng Li}
\affiliation{School of Engineering, Westlake University, Hangzhou, China}
\affiliation{Institute of Advanced Technology, Westlake Institute for Advanced Study, Hangzhou, China}

\author{Junfang Zhao}%
\affiliation{School of Science, China University of Geosciences (Beijing), Beijing, China}

\author{Hui Xiang \textsuperscript{*}}%
\affiliation{Science42 Technology, Beijing, China}%
\email{xianghui@science42.tech}

\author{Dixia Fan \textsuperscript{*}}
\email{fandixia@westlake.edu.cn}
\affiliation{School of Engineering, Westlake University, Hangzhou, China}
\affiliation{Institute of Advanced Technology, Westlake Institute for Advanced Study, Hangzhou, China}

\begin{abstract}
Physics-informed neural networks (PINNs) have emerged as a transformative framework for addressing operator learning and inverse problems involving the Korteweg-de Vries (KdV) equation for internal solitary waves. By integrating physical constraints with data-driven optimization, PINNs overcome the critical challenges of parameter unmeasurability in the KdV equation for internal solitary waves in two-layer fluid systems.
This work addresses two problems: (1) Operator learning constructs a mapping from parameters to solutions, enabling wave evolution predictions from unknown parameters. Comparative studies demonstrate prediction errors as low as $10^{-4}$ when using 1000 training points. (2) Inverse problem solving leverages sparse and potentially noisy observational data with physics-regularized constraints to invert nonlinear coefficients successfully.
Compared to conventional approaches, this end-to-end differentiable paradigm unifies operator learning and inverse problem-solving while overcoming mesh discretization errors and high-dimensional parameter space iteration costs.
The method shows effectiveness for internal wave problems in stratified fluids, providing both accurate forward modeling and robust parameter inversion capabilities, even under noise.
\end{abstract}

\maketitle

\section{\label{sec:level1}INTRODUCTION}
Internal solitary waves, frequently observed in density stratified oceans, exert substantial influence on the safety of marine structures \citep{Zhang2023,Cheng2024,Wang2024}, underwater acoustic propagation \citep{Li2023,Zheng2024}, and mass transport processes \citep{Brandt2014,Deepwell2016}. These waves have emerged as a research hotspot within both the mathematical and engineering disciplines. When the density of seawater undergoes a rapid transition resulting in a relatively narrow pycnocline the system can be effectively modeled as a two-layer fluid \citep{Helfrich2006,Choi2022}. Among the mathematical models used to describe internal solitary waves in such systems, the Korteweg-de Vries (KdV) equation \citep{Benjamin1966} is particularly prominent because of its relatively simple but effective formulation. Extensive studies have shown that the KdV equation accurately captures the fundamental characteristics of small-amplitude internal solitary waves, including wave profiles, propagation speeds, and velocity fields \citep{Grue1999, Kodaira2016, Xuan2024}.

Traditionally, the KdV equation has been approached through various analytical and numerical methods. Analytically, the inverse scattering transform (IST) offers a robust technique for addressing initial value problems by converting the nonlinear KdV equation into a linear framework, thereby facilitating the exact construction of soliton solutions and enabling the exploration of their interactions. Perturbation methods have also been applied to approximate solutions for small-amplitude waves by assuming weak nonlinearity and dispersion. On the numerical front, finite difference methods (FDM) and spectral methods are commonly adopted to discretize both spatial and temporal domains, thereby simulating the evolution of wave propagation. These numerical strategies prove especially beneficial when examining the dynamics of internal solitary waves under complex initial and boundary conditions.

Despite their success, traditional solution methods for the KdV equation present several limitations. The IST method, while exact, is computationally demanding and restricted to specific initial conditions that yield soliton solutions, making it impractical for handling general or noisy data. Perturbation methods, on the other hand, are constrained to weakly nonlinear regimes and fail to capture large-amplitude waves or strongly nonlinear interactions. Numerical methods such as FDM and spectral methods require careful discretization of the computational domain, which may lead to numerical instabilities or dispersion errors, particularly in long-term simulations or in scenarios with steep gradients. Furthermore, these conventional approaches often impose significant computational burdens when dealing with high-dimensional problems or multiscale phenomena, such as interactions between internal solitary waves and varying bathymetry or stratification.

Recent breakthroughs in deep learning have introduced scientific machine learning as a transformative paradigm in computational science. In this context, Physics-Informed Neural Networks (PINNs) \cite{cuomoScientificMachineLearning2022} have been developed as an innovative hybrid modeling framework that seamlessly integrates data-driven models with traditional physics-based numerical methods through the incorporation of hard constraints imposed by differential equations. 
A series of pioneering studies by Raissi, Perdikaris, and Karniadakis \cite{raissiPhysicsInformedDeep2017,raissiPhysicsInformedDeep2017a,raissiPhysicsinformedNeuralNetworks2019} systematically incorporated governing equations and initial/boundary conditions of partial differential equations (PDEs) into neural network architectures using residual formulations, thereby establishing an end-to-end PDE-solving framework leveraging automatic differentiation. PINNs have demonstrated remarkable versatility in addressing both forward and inverse problems across diverse scientific disciplines. 
This integration established an end-to-end framework for solving PDEs by leveraging automatic differentiation. PINNs have since demonstrated considerable versatility in addressing both forward and inverse problems across a broad spectrum of disciplines. Early applications validated their effectiveness in wave equation modeling \cite{moseleySolvingWaveEquation2020}, parameter estimation \cite{caiPhysicsInformedNeuralNetworks2021a}, and streamlined implementations \cite{bafghiPINNsTF2FastUserFriendly2023}, while subsequent studies expanded their applicability to multiphysics problems \cite{kadeethumPhysicsInformedNeuralNetworks2020}. Their capacity for nonlinear function approximation enables them to naturally resolve multiscale phenomena, discontinuities, and strong nonlinearities, as evidenced by their success in high-dimensional fluid dynamics \cite{caiPhysicsinformedNeuralNetworks2021}, quantum mechanical parameter inference \cite{jinPhysicsInformedNeuralNetworks2022}, and materials microstructure analysis \cite{zhangAnalysesInternalStructures2022}. Furthermore, hardware acceleration through parallel computing techniques specifically tailored for PINNs \cite{escapil-inchauspeHAnalysisDataparallelPhysicsinformed2023} has significantly reduced training times, thereby rendering large-scale simulations feasible. PINNs effectively overcome many of the limitations associated with traditional methods, offering a robust framework for tackling complex scientific challenges.

Recent progress in PINNs has further established a unified framework for simultaneously solving inverse problems and learning solution operators for nonlinear PDEs. By integrating physics-constrained architectures with data-efficient optimization strategies, PINNs enable (1) high-fidelity parameter inference and (2) continuous operator learning \cite{Lu_2021,li2021fourierneuraloperatorparametric}. 
This dual capability not only regularizes traditionally ill-posed inverse problems through PDE-constrained optimization but also establishes differentiable mappings from input parameters to field solutions—a development that is particularly impactful for applications involving the KdV equation. Given the widespread application of the KdV equation in modeling phenomena across water waves, plasma physics, and elastic media—where key parameters are often not directly measurable in field experiments—solving the inverse problem via PINNs offers a promising pathway forward. Additionally, a PINN-based operator facilitates the mapping between parameter spaces and corresponding solution spaces in KdV simulations.
Motivated by these considerations, this study investigates both the operator learning problem and the inverse problem for the KdV equation, demonstrating the advantages of PINNs in addressing these challenges.

The remainder of this paper is structured as follows. Section~\ref{Methodology} introduces the KdV equation and provides a foundational overview of PINNs. Section~\ref{Learning the solution operator} evaluates the performance of PINNs in learning solution operators for the KdV equation. Section~\ref{Inverse Problem} focuses on solving the inverse problem, examining the effects of parameter selection, data sparsity, measurement noise, and measurement regions. Finally, Section~\ref{Conclusion} presents conclusions and potential directions for future research.

\section{Methodology}\label{Methodology}
\subsection{The KdV Equation}
\begin{figure*}[t]
\includegraphics[width=\textwidth]{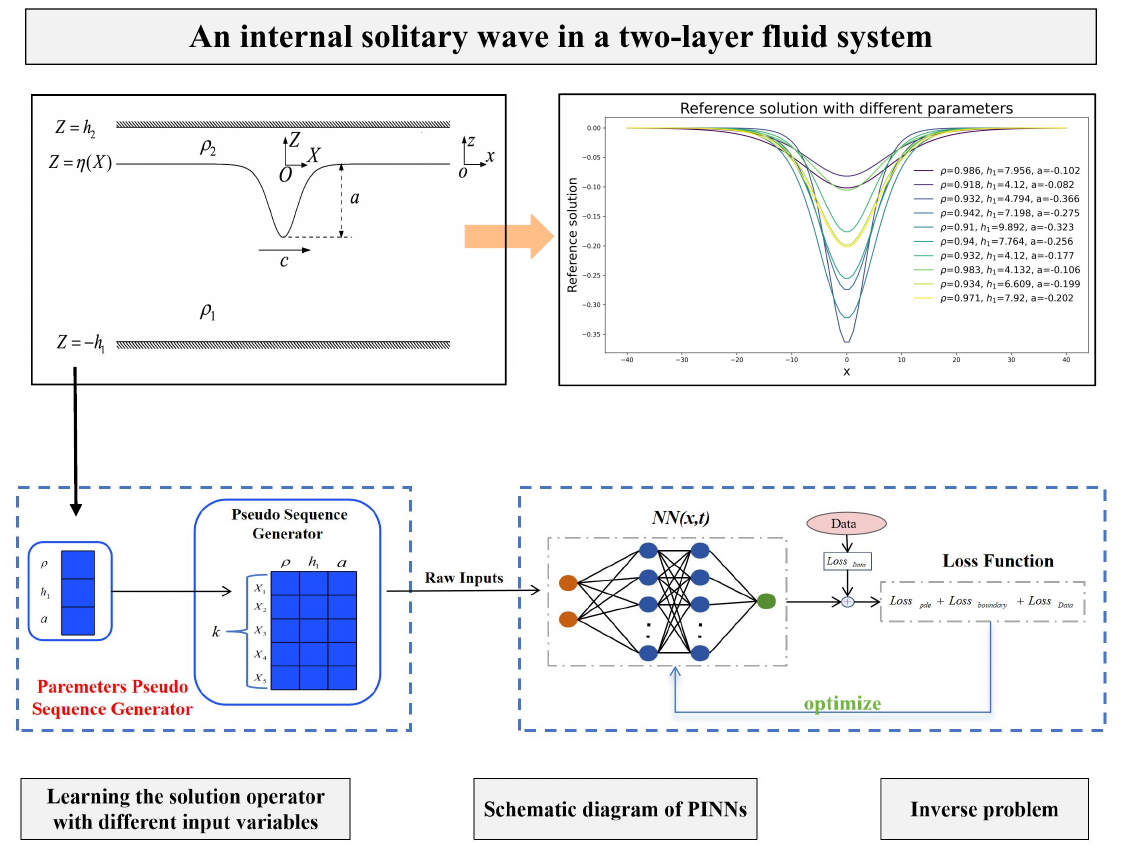}
\caption{\label{diagram} Schematic diagram of solving an internal solitary wave in a two-layer fluid system by PINNs.}
\end{figure*}

Consider a system consisting of two fluid layers with densities and thicknesses given by $\rho_i$ and $h_i$, respectively, where $i = 1$ denotes the lower fluid layer and $i = 2$ denotes the upper fluid layer.
In two dimensions, $x$ is the horizontal coordinate (positive to the right) and $z$ is the vertical coordinate (positive up), and the origin of the coordinate is located at the undisturbed interface between the two fluid layers. 
The upper surface is considered a rigid lid, expressed as $z = h_2$, where $h_2$ is the undisturbed and constant depth of the upper fluid layer. 
The interface between the two fluid layers and the seabed is expressed as $z = \eta(x,t)$ and $z = -h_1$, respectively, where $h_1$ is the undisturbed and constant depth of the lower fluid layer.
The schematic diagram of the two-layer fluid system is shown in Fig.~\ref{diagram}.

The KdV equation can be written as follows \citep{Kodaira2016}: 
\begin{equation}\label{KdV}
    \begin{aligned}
        \frac{\partial \eta}{\partial t}+c_0\frac{\partial \eta}{\partial x}+c_1\eta\frac{\partial \eta}{\partial x}+c_2\frac{\partial ^3 \eta}{\partial x^3}=0,
    \end{aligned}
\end{equation}
where 
\begin{equation}
    \begin{aligned}
        c_0&=\sqrt{\frac{gh_1h_2(\rho_1-\rho_2)}{\rho_1h_2+\rho_2h_1}}, \\
        c_1&=-\frac{3c_0}{2}\frac{\rho_2h_1^2-\rho_1h_2^2}{\rho_2h_2h_1^2+\rho_1h_1h_2^2},\\
        c_2&=\frac{c_0}{6}\frac{\rho_2h_2^2h_1+\rho_1h_2^2h_2}{\rho_2h_1+\rho_1h_2},
    \end{aligned}
\end{equation}
$\eta(x,t)$ is the wave elevation, $t$ is the time and $g$ is the gravitational acceleration.

Introducing a wave-following coordinate $X=x-ct$ and $Z=z$, where $c$ is the internal wave speed, Eq. (\ref{KdV}) can be derived in the following form :
\begin{equation}\label{KdV_X}
    \begin{aligned}
        (-c+c_0+c_1\eta)\frac{\text{d} \eta}{\text{d} X}+c_2\frac{\text{d} ^3 \eta}{\text{d} X^3}=0,
    \end{aligned}
\end{equation}
with the boundary conditions $\frac{\text{d} \eta}{\text{d} X}=0$ when $X=0$ and $\eta=0$ when $X\to \infty$. 

Eq. (\ref{KdV_X})  has an exact solution:
\begin{equation}\label{exact-solution}
    \begin{aligned}
      \eta(X)=a\text{sech}^2(X/\lambda),
    \end{aligned}
\end{equation}
with parameters
\begin{equation}
    \begin{aligned}
      \lambda&=\sqrt{\frac{12c_2}{ac_1}},\\
      c&=c_0+\frac{ac_1}{3},
    \end{aligned}
\end{equation}
and $a$ is the amplitude of the internal solitary wave.

In the present study, we consider the cases for small-amplitude internal solitary waves in a two-layer fluid system where the density ratio between the two-fluid layers is close to $1$.

The parameters of the cases range from
\begin{equation}\label{parameters}
    \begin{aligned}
        \frac{\rho_2}{\rho_1}&\in [0.9,1],\\
        \frac{h_2}{h_1}&\in [\frac{1}{10},\frac{1}{4}],\\
        \frac{a}{h_2}&\in [-0.4,-0.05],
    \end{aligned}
\end{equation}
where the undisturbed constant depth of the upper-fluid layer is $h_2=1$m. 
It should be noted that the KdV equation under the rigid-lid assumption can describe the internal solitary wave in a two-layer fluid system well under these parameters \citep{Grue1999,Kodaira2016}.
\subsection{Physics-Informed Neural Networks (PINNs)}
In this section, we briefly introduce the basics of PINNs, illustrated for the case of \eqref{KdV}. PINNs represent a novel computational approach that integrates physical laws with deep learning. Building upon traditional neural networks, PINNs incorporate the residuals of physical equations as constraints into the loss function, enabling effective modeling of physical systems. 
 Specifically, a fully connected neural network $\hat{\eta}(x, t; \theta)$ with parameters $\theta$ (weights and biases) is constructed to approximate the solution $\eta(x, t)$. In this work, the network consists of 4 hidden layers with 20 neurons per layer and utilizes the hyperbolic tangent (tanh) activation function. The derivatives required to compute the PDE residual are obtained via automatic differentiation (AD), a key component of modern deep learning frameworks that provides exact derivatives of the network output with respect to its inputs and parameters.

As illustrated schematic diagram of PINNs in Fig.~\ref{diagram}, we construct a neural network to solve the KdV Eq. \eqref{KdV} with inputs $(x,t)$, and outputs $\hat{\eta}_1(x,t)$. The loss function of PINNs consists of three components: PDE Residual Loss ($Loss_{pde}$), Boundary Condition Loss ($Loss_{bc}$), and Data Loss ($Loss_{data}$).
The PDE Residual Loss quantifies the deviation of the neural network's predictions from the governing physical laws, typically measured by computing the residuals of the governing equations at a set of collocation points $\{ (x_r^i, t_r^i) \}_{i=1}^{N_r}$ within the domain. The Boundary Condition Loss enforces the prescribed conditions at the domain boundaries $\{ (x_{bc}^i, t_{bc}^i) \}_{i=1}^{N_{bc}}$.
The Data Loss measures the discrepancy between the neural network's predictions and the actual observed data at measurement points $\{ (x_d^i, t_d^i) \}_{i=1}^{N_d}$. These components are combined into a total loss function through a weighted sum:
\begin{equation} \label{eq:loss}
    \begin{aligned}
        Loss_{total}  = & w_{pde}Loss_{pde}+w_{bc}Loss_{bc}+w_{data}Loss_{data},\\
        Loss_{pde}  = \frac{1}{N_r} &\sum_{i=1}^{N_{r}}\|\frac{\partial \hat{\eta}(x_r^i,t_r^i)}{\partial t}+c_0\frac{\partial \hat{\eta}}{\partial x}+c_1\hat{\eta}(x_r^i,t_r^i)\frac{\partial \hat{\eta}(x_r^i,t_r^i)}{\partial x} \\
        &+c_2\frac{\partial ^3 \hat{\eta}(x_r^i,t_r^i)}{\partial x^3}\|^2,\\
        Loss_{bc}  = \frac{1}{N_{bc}} & \sum_{i=1}^{N_{bc}}\|\hat{\eta}(x_{bc}^i,t_{bc}^i)\|^2, \ \text{for } |x_{bc}^i| \to \infty, \\
        Loss_{data}  = \frac{1}{N_d} & \sum_{i=1}^{N_d}\|\hat{\eta}(x_{d}^i,t_{d}^i)-\eta_{obs}(x_{d}^i,t_{d}^i)\|^2,
    \end{aligned}
\end{equation}
where $w_{pde},w_{bc},w_{data}$ denote the weighting coefficients for the Physics Loss, boundary Loss and Data Loss, respectively. In this study, we used constant weights $w_{pde}=0.1$, $w_{bc}=1$, and $w_{data}=10$, prioritizing fitting the observational data while ensuring physical consistency. While adaptive weighting schemes exist to mitigate potential gradient pathologies or imbalances between loss terms, fixed weights proved sufficient for the problems considered here. $N_{r}, N_{bc}, N_{d}$ denote the number of residual sampling points, boundary sampling points, and data sampling points, respectively. Note that $\eta_{obs}$ represents the observed data, which may be exact synthetic data or noisy data.

By minimizing this loss function using optimization algorithms (such as Adam or L-BFGS), 
the network parameters $\theta$ (and potentially unknown physical parameters) are adjusted so that the solution can simultaneously satisfy physical constraints and accurately fit the observed data.
PINNs are not only effective for solving forward problems (i.e., predicting solutions given known physical laws and boundary conditions), but also particularly well-suited for inverse problems. Inverse problems typically involve inferring unknown parameters, initial conditions, or boundary conditions of a model from observed data. The core advantage of PINNs in inverse problems lies in their ability to incorporate physical laws as prior knowledge into the learning procedure, thereby improving the accuracy and robustness of parameter estimation even when observational data is limited or noisy.

In the context of inverse problems, PINNs treat unknown parameters (e.g., $c_0, c_1, c_2$ or $\rho_1, h_1$) as trainable variables alongside the neural network parameters $\theta$. The optimization procedure minimizes the total loss function with respect to both $\theta$ and the unknown physical parameters. Specifically, the loss function encompasses not only the Data Loss but also the Physics Loss and boundary/initial condition losses, ensuring that the estimated parameters are consistent with both the observed data and the underlying physical laws.

For example, consider the KdV Eq. \eqref{KdV} with unknown coefficients \(c_0\), \(c_1\), \(c_2\) or underlying physical parameters like \(\rho_1\), \(\rho_2\), $h_1$. PINNs can estimate these coefficients using observational data \(\eta_{obs}(x_d^i, t_d^i)\). During this procedure, the Physics Loss \(Loss_{pde}\) incorporates dependencies on the unknown coefficients, and the optimization procedure simultaneously adjusts both the neural network parameters $\theta$ and the unknown physical parameters to minimize the total loss function.

Applying PINNs to inverse problems provides a flexible and efficient approach for addressing challenges such as parameter estimation and model calibration, particularly in scenarios with sparse data or moderate noise levels. The impact of noise is further investigated in Sec.~\ref{sec:noise_impact}.

\section{Numerical Experiments and Results}
 All numerical experiments were performed using the Pytorch framework on a standard desktop computer equipped with an NVIDIA RTX 6000 Ada Generation. The neural network was trained using the Adam optimizer for training until convergence criteria were met, with 20,000 iterations for the operator learning problem and 25,000 iterations for the inverse problem. Collocation points for the PDE residual were sampled uniformly across the spatiotemporal domain $x \in [-40, 40], t \in [0, 4]$, while boundary points were sampled at the spatial boundaries $x=\pm 40$ for all $t$.

\subsection{Learning the Solution Operator}\label{Learning the solution operator}

Physics-Informed Neural Networks (PINNs) provide a powerful framework for solving problems where the governing equation parameters are unknown, enabling the direct inference of solutions. In this section, we focus on learning solution operators for systems with unknown parameters. Specifically, we aim to learn the mapping $\mathcal{G}: \mathbf{s} \to \eta(X)$, which associates the unknown parameters $\mathbf{s}$ with the corresponding solution $\eta(X)$ of Eq. \eqref{KdV_X}.

By analyzing the parametric form of Eqs. \eqref{KdV_X} and \eqref{parameters}, we identify the unknown parameters as $\mathbf{s}=(\frac{\rho_2}{\rho_1}, h_1, a)$. Defining $\rho=\frac{\rho_2}{\rho_1}$, the problem reduces to learning the solution operator $\mathcal{G}: (\rho, h_1, a) \to \eta(X)$. We construct a neural network similar to the one described in Sec. II.B, where the input consists of the unknown parameters $\mathbf{s}=(\rho, h_1, a)$ along with the spatial coordinate $X,$ and the output is the predicted solution $\hat{\eta}(X; \mathbf{s})$.
 
Moreover, since the solution $\eta(X)$ inherently depends on the spatial variable $X$ for any given set of parameters, we introduce an additional component between the input layer and the neural network to explicitly learn and regularize spatial dependencies. As illustrated in the left part of Fig.~\ref{diagram}, this approach expands the unknown parameter inputs into a pseudo-sequential sequence that reinforces the input states with spatial dependencies at $X_i$. This pseudo-sequential sequence is then processed by the PINN framework, yielding a sequentially predicted solution $\hat{\eta}(X_i)$ corresponding to the given parameters. This enhancement ensures a more structured and spatially aware learning process, improving the accuracy of the learned solution operator.

 The network learns to predict the solution $\eta$ at any given $X$ for the input parameter set $(\rho, h_1, a)$. The training data consists of pairs of parameter sets and corresponding exact solutions $\eta(X)$ sampled at various points $X$. The parameter sets are sampled from the ranges defined in Eq.~\eqref{parameters} using Uniform sampling. For each parameter set, the exact solution $\eta(X)$ is computed by Eq. \eqref{exact-solution} and sampled at $N_X=3000$ points uniformly.

To evaluate the effectiveness of the proposed model, we conducted systematic experiments under varying training data regimes (number of parameter sets $(\rho, h_1, a)$). As shown in Fig.~\ref{operator-fig}, our results reveal a strong correlation between data availability and solution accuracy. This comparative analysis highlights the critical role of training data quantity in ensuring precise and reliable solutions. Furthermore, the quantitative error analysis presented in Table \ref{operator-table} substantiates this observation. Key performance metrics, such as the mean absolute error (MAE), exhibit consistent improvement with larger training datasets, underscoring the direct relationship between data volume and solution precision. These findings emphasize the necessity of sufficient training data in enhancing the accuracy and robustness of PINNs for parameter estimation and function approximation tasks.
 
\begin{figure*}[t]
\includegraphics[width=\textwidth]{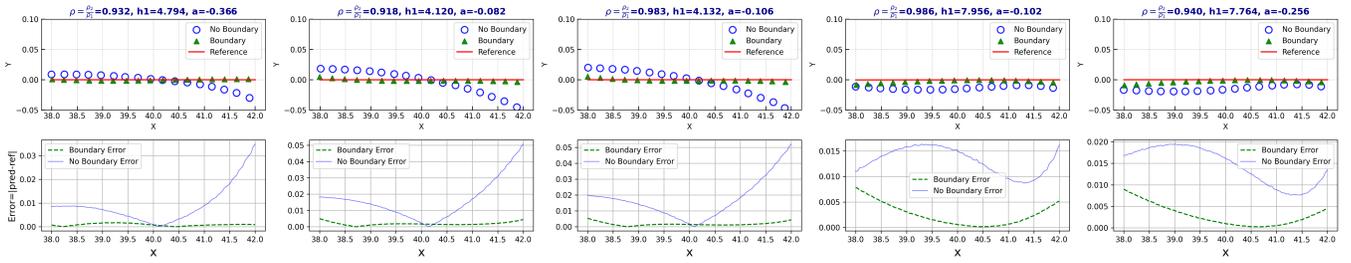}
\caption{\label{compare-boundary}Impact of the boundary condition loss on the PINN-predicted solution.}
\end{figure*}

\begin{figure*}[t]
  \centering
  \includegraphics[
  trim=0mm 20mm 0mm 0mm, 
  clip, 
  width=0.85\textwidth]{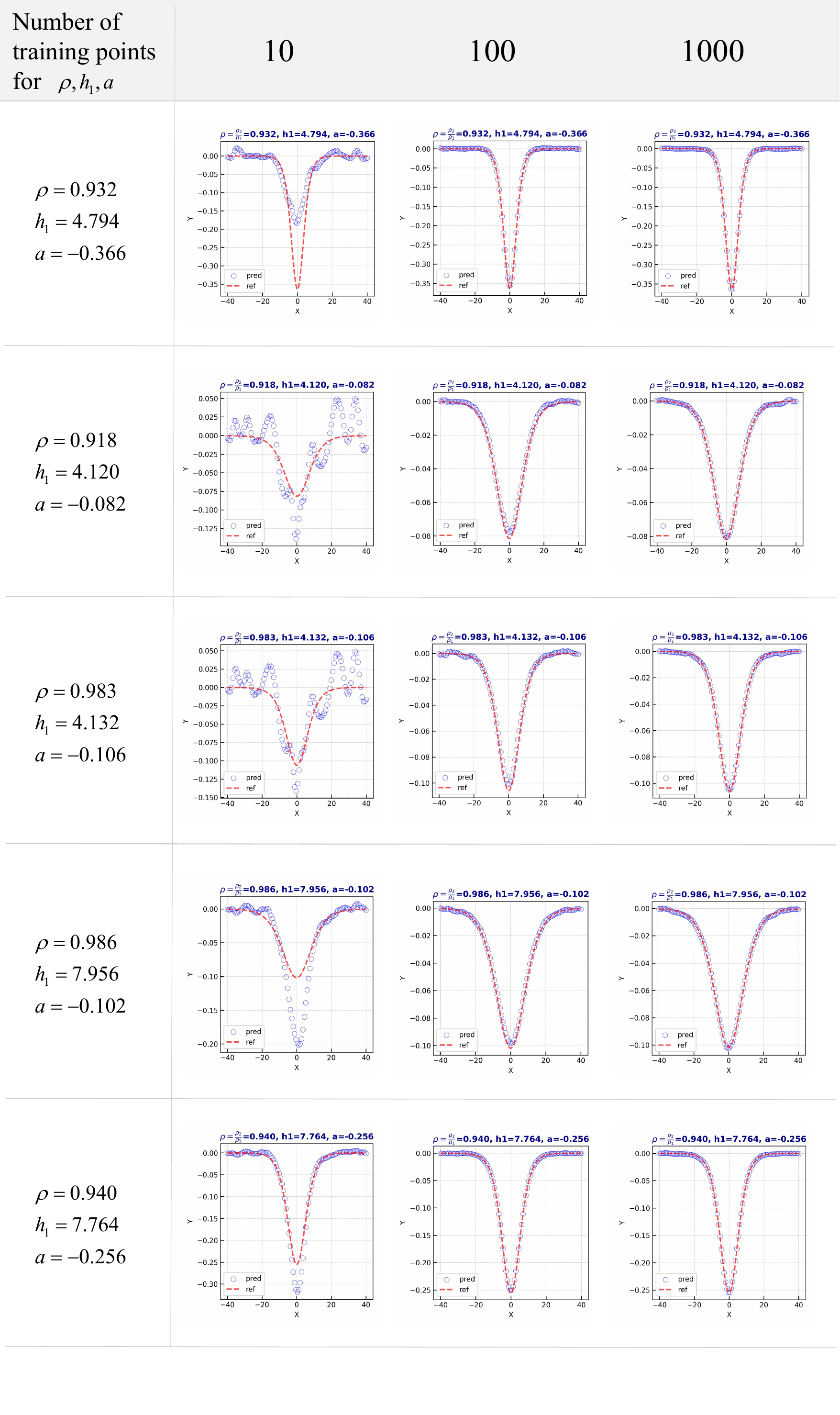} 
  \vspace{-\baselineskip}
  \caption{Learning the solution operator: A comparison between the exact solution and the predicted solution $\hat{\eta}(X)$ for different input variables $(\rho,h_1,a)$. Results are shown for models trained with varying numbers of training parameter sets.}
  \label{operator-fig}
\end{figure*}

To specifically illustrate the impact of boundary error, we trained models both with and without the inclusion of $Loss_{bc}$ (i.e., setting $w_{bc}=0$) to solve this problem. All other settings, including the network architecture, training data points (if applicable), and the weights for $Loss_{pde}$ and $Loss_{data}$, were kept identical.
It is clearly observable from Fig.~\ref{compare-boundary} that when $Loss_{bc}$ is included in the loss function, the PINN learns the correct far-field asymptotic behavior. Its predicted solution stably approaches zero for large values of $X$, including in regions intentionally examined outside the training domain via extrapolation. This aligns perfectly with the physical characteristics of solitary waves.
In contrast, when $Loss_{bc}$ is removed from the loss function, although the PINN might still provide a seemingly reasonable solution within the region containing training data points or where PDE residual constraints are strong, its behavior becomes unreliable when extrapolating towards the far-field especially outside the training domain. Specifically, the solution does not decay to zero as physically expected with increasing $X$; instead, it may diverge or maintain a non-zero level. This clearly demonstrates that explicitly imposing the boundary condition loss is indispensable for constraining the neural network to produce physically intuitive solutions in regions where data is sparse or direct PDE constraints are weak, such as far-field boundaries. Without this term, the PINN's predictive behavior in these regions is primarily governed by the network's own extrapolation characteristics, which can easily lead to non-physical results and, consequently, large boundary errors.
Therefore, in all subsequent numerical experiments presented in this work, the boundary condition $Los_{bc}$ is incorporated as a standard component of the total PINN loss function to ensure physically reliable and globally accurate solutions.

\begin{table*}[t]
\centering
\renewcommand\arraystretch{1.2}
\begin{tabular}{ >{\raggedright\arraybackslash}p{0.25\linewidth}  >{\centering\arraybackslash}p{0.2\linewidth} >{\centering\arraybackslash}p{0.2\linewidth} >{\centering\arraybackslash}p{0.2\linewidth} }
\toprule
\multirow{3}{*}{\textbf{\centering Test parameter set}~$(\rho, h_1, a)$} 
& \multicolumn{3}{c}{\textbf{Mean Absolute Error (MAE)}}  \\
\cmidrule(lr){2-4}
& \multicolumn{3}{c}{\textbf{Number of Training Parameter Sets}} \\
\cmidrule(lr){2-4}
 & \textbf{10} & \textbf{100} & \textbf{1000} \\
\midrule
$(0.932, 4.794, -0.366)$
& 2.195e-02
& 1.548e-03
& 4.425e-04 \\
\addlinespace
$(0.918, 4.120, -0.082)$
& 1.950e-02
& 1.377e-03
& 5.873e-04 \\
\addlinespace
$(0.983, 4.132, -0.106)$
& 1.792e-02
& 1.734e-03
& 7.517e-04 \\
\addlinespace
$(0.986, 7.956, -0.102)$
& 1.574e-02
& 1.237e-03
& 5.256e-04 \\
\addlinespace
$(0.940, 7.764, -0.256)$
& 7.831e-02
& 6.574e-03
& 4.900e-04 \\
\bottomrule
\end{tabular}
\caption{The Mean Absolute Error (MAE = $\frac{1}{N_X}\sum_i |\hat{\eta}(X_i) - \eta(X_i)|$) between the exact solution and predicted solution of Eq.~\eqref{KdV_X}, evaluated on test parameter sets $(\rho, h_1, a)$, using models trained with different numbers of training parameter sets.}
\label{operator-table}
\end{table*}

\subsection{Inverse Problem: Parameter Estimation}\label{Inverse Problem} 

PINNs provide a powerful framework for solving inverse problems by integrating observational data with governing physical laws. For parameter estimation in the KdV equation, the PINN approach involves simultaneously optimizing the neural network parameters (weights and biases) and the unknown physical parameters by minimizing a composite loss function. In this section, unless otherwise specified (see Sec.~\ref{sec:noise_impact}), the "observational data" ($\eta_{obs}$) consists of synthetic data generated by sampling the exact analytical solution $\eta(x,t)$ corresponding to known ground truth parameters at specified spatiotemporal points $(x_d^i, t_d^i)$. Noise is not added to this synthetic data in Sections III.B.1-III.B.3. This loss function typically comprises three components as defined in Eq.~\eqref{eq:loss}:
\begin{itemize}
    \item \textbf{Data Fidelity Loss}: Ensures consistency between the network’s predictions $\hat{\eta}(x_d^i, t_d^i)$ and sparse observational data points $\eta_{obs}(x_d^i, t_d^i)$.
    \item \textbf{Physics-Informed Loss}: Enforces the satisfaction of the KdV equation \eqref{KdV}, typically by penalizing its residual across collocation points in the spatiotemporal domain.
    \item \textbf{Boundary/Initial Condition Loss}: Ensures adherence to the prescribed constraints at the boundaries or initial time.
\end{itemize}

The inverse problem is thus formulated as an optimization task. Gradient-based algorithms (e.g., Adam or L-BFGS) are employed to iteratively adjust both the network weights and the target physical parameters to minimize the total loss. This dual optimization mechanism enables PINNs to infer unknown parameters, often without requiring exhaustive measurements or strong prior knowledge of the parameter distributions.

\subsubsection{Parameter Selection and Estimation Accuracy}\label{sec:param_select}
In practical two-layer fluid systems, the total height $h_1+h_2$ is often measurable through remote sensing or direct observation, leaving the height ratio $h_1/h_2$ as the critical unknown. Similarly, while the upper-layer density $\rho_2$ might be more accessible via in-situ instruments, the lower-layer density $\rho_1$ remains challenging to measure due to depth limitations. We therefore prioritize the estimation of $h_1/h_2$ and $\rho_1$, acknowledging their significance in characterizing interfacial wave dynamics. 
To evaluate PINN’s capacity for multi-parameter recovery, we systematically test three scenarios (Fig.~\ref{Fig.1}), where the goal is to recover the true parameters $h_1/h_2 = 4.13$, $\rho_1 = 0.977$, $\rho_2 = 1.0$ (using $h_2=1$, $a=-0.77$, $g=9.81 m/s^2$ to generate the exact solution data $\eta(x,t)$). The relative error used in the figures is defined as: 
$$\text{Relative Error} = \frac{|\text{Predicted Value} -  \text{True Value}|}{|\text{True Value}|}.$$ 
The observational data for these cases uses $N_x=256$ spatial points and $N_t=201$ temporal points ($N_d = 51456$).:

\begin{figure*}[t]
\centering
    \subfloat[Estimate $h_1/h_2, \rho_1, \rho_2$\label{fig1:a}]{
        \includegraphics[width=0.3\linewidth]{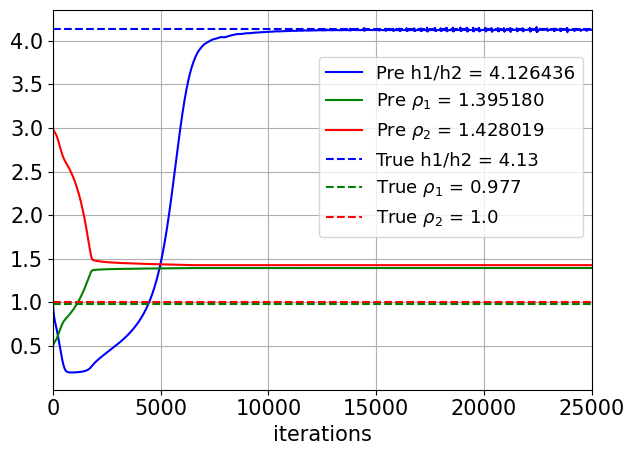}
    }
    \subfloat[Estimate $h_1/h_2, \rho_1$ (fix $\rho_2$)\label{fig1:b}]{
        \includegraphics[width=0.3\linewidth]{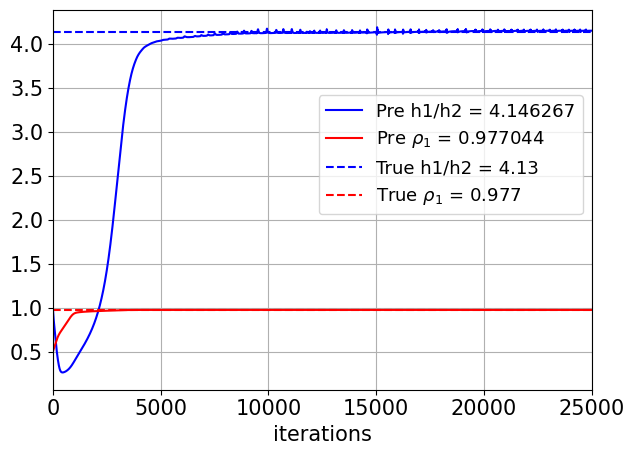}
    }
    \subfloat[Estimate $h_1/h_2$ (fix $\rho_1, \rho_2$)\label{fig1:c}]{
        \includegraphics[width=0.3\linewidth]{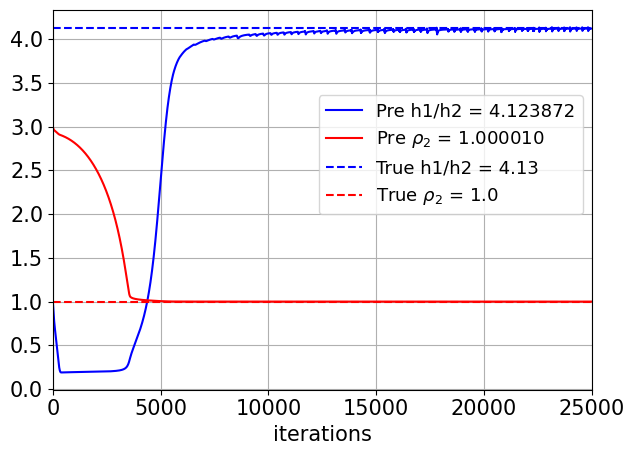}
    }
    \caption{Convergence of parameter estimation over 25,000 training iterations for different scenarios using dense data ($N_x=256, N_t=201$). Dashed lines represent true values, solid lines represent PINN predictions. (a) Simultaneous estimation of $h_1/h_2$, $\rho_1$, and $\rho_2$. (b) Estimation of $h_1/h_2$ and $\rho_1$ with $\rho_2$ fixed. (c) Estimation of $h_1/h_2$ with both $\rho_1$ and $\rho_2$ fixed.}
    \label{Fig.1}
\end{figure*}

\begin{figure*}[t]
\centering
    \subfloat[Exact Solution $\eta(x,t)$\label{fig2:a}]{
        \includegraphics[width=0.3\linewidth]{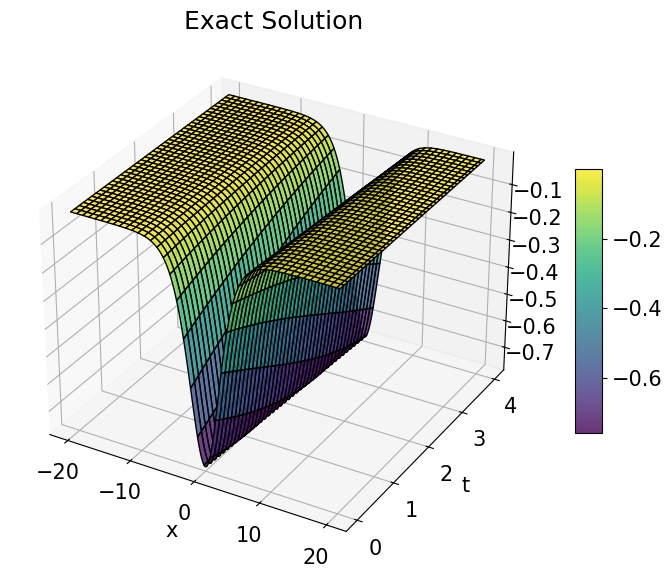}
    }
    \subfloat[Predicted Solution $\hat{\eta}(x,t)$\label{fig2:b}]{
        \includegraphics[width=0.3\linewidth]{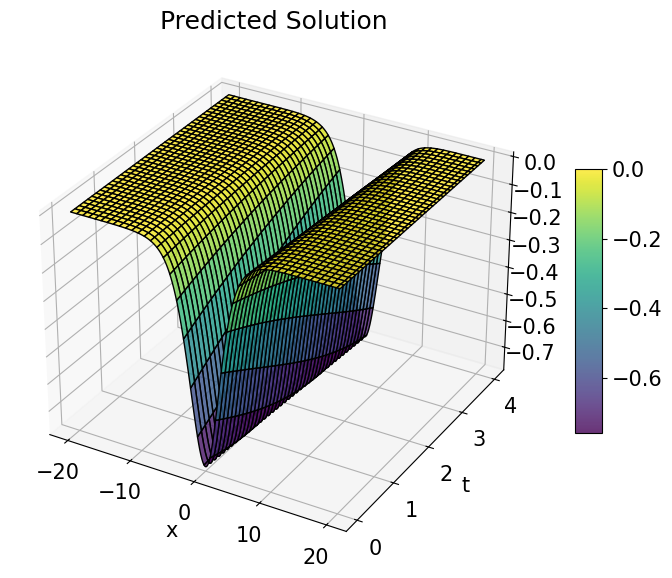}
    }
    \subfloat[Absolute Error $|\eta - \hat{\eta}|$\label{fig2:c}]{
        \includegraphics[width=0.3\linewidth]{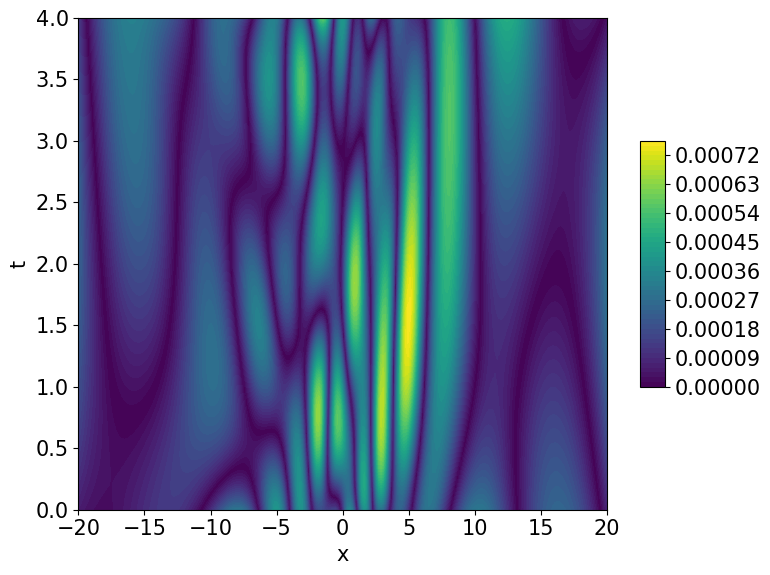}
    }
    \caption{Comparison of final field results for the parameter estimation of $h_1/h_2$ and $\rho_1$ (Case (b) from Fig.~\ref{fig1:b}). (a) Exact solution, (b) PINN predicted solution, and (c) Absolute error between exact and predicted solutions.}
    \label{Fig.2}
\end{figure*}

\begin{itemize}
\item[$\bullet$]\textbf{Case (a)}:  Simultaneous estimation of $h_1 /h_2, \rho_1$, and $\rho_2$ yields unsatisfactory results (relative errors $> 15\%$). This difficulty arises because the KdV coefficients ($c_0, c_1, c_2$) depend nonlinearly on these parameters in a coupled manner. Variations in one parameter can potentially be compensated by variations in others, leading to similar wave dynamics ($\eta(x,t)$). This weak identifiability makes it challenging to uniquely determine all three parameters from limited wave observations. From a PINN perspective, this parameter coupling can lead to a complex, non-convex loss landscape with flat regions or multiple local minima, hindering the optimizer's ability to converge to the true global minimum corresponding to the correct parameter values. This highlights inherent limitations in the KdV equation’s parameter sensitivity when multiple unknowns coexist.
\item[$\bullet$]\textbf{Case (b)}: Fixing $\rho_2$ while estimating $h_1 /h_2 $ and $\rho_1$ achieves sub-1\% relative errors, demonstrating precise recovery of physically interdependent parameters, when the number of unknown, strongly coupled parameters is reduced. 
\item[$\bullet$]\textbf{Case (c)}: Estimating $h_1 /h_2$ alone with known densities produces near-perfect accuracy (relative error $< 0.1\%$), confirming the method’s reliability for single-parameter inversion.
\end{itemize}
Fig.~\ref{Fig.2} visually corroborates these findings through side-by-side comparisons of ground truth solutions, PINN predictions, and absolute error distributions for Case (b). 
The excellent agreement (max absolute error $< 10^{-3}$) in both height ratio and density estimations validates the physical consistency of the PINN-inferred parameters ($h_1/h_2$ and $\rho_1$).

\subsubsection{Data Efficiency and Spatial Sparsity}\label{sec:sparsity}
\begin{figure*}[t]
    \centering
    \subfloat[Relative error vs. $N_x$\label{fig3:a}]{
        \includegraphics[width=0.48\textwidth]{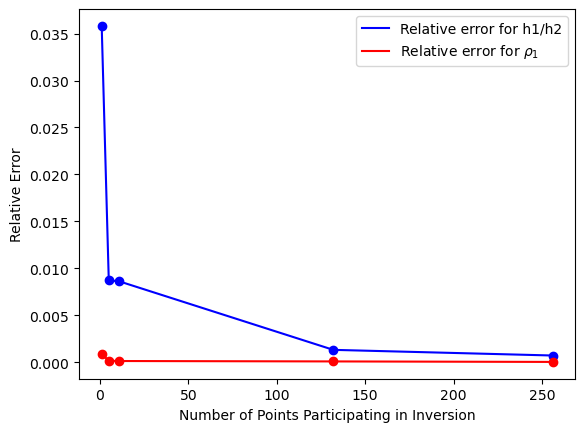}
    }
    \hfill
    \begin{minipage}[b]{0.48\textwidth}
        \centering
        \subfloat[$N_x=256$\label{fig3:b}]{
            \includegraphics[width=0.48\linewidth]{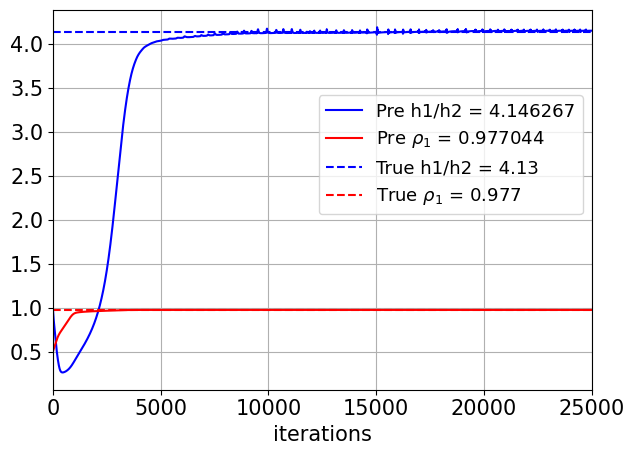}
        }
        \hfill
        \subfloat[$N_x=11$\label{fig3:c}]{
            \includegraphics[width=0.48\linewidth]{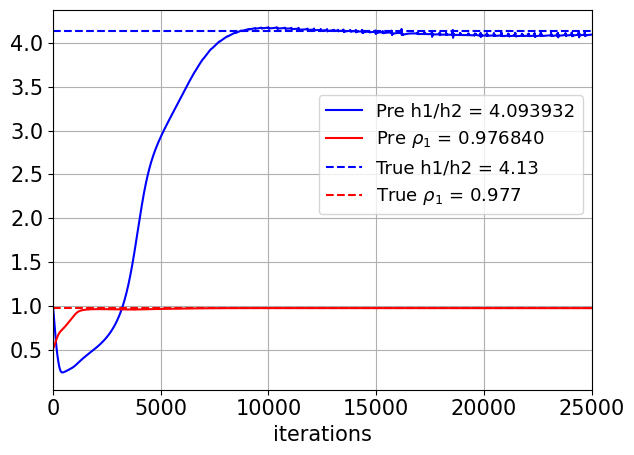}
        }
        \vfill
        \subfloat[$N_x=5$\label{fig3:d}]{
            \includegraphics[width=0.48\linewidth]{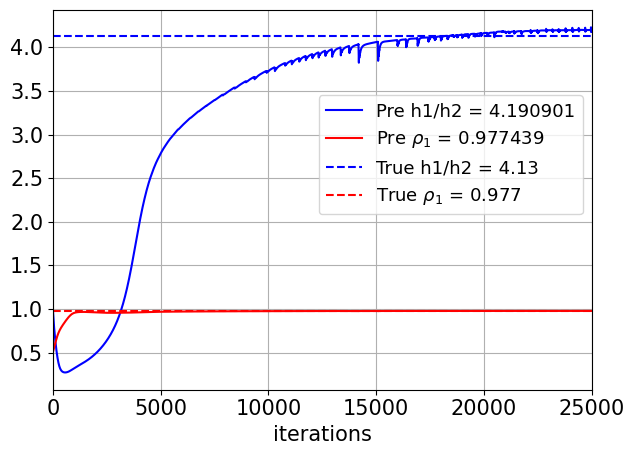}
        }
        \hfill
        \subfloat[$N_x=1$ (at $x=0$)\label{fig3:e}]{
            \includegraphics[width=0.48\linewidth]{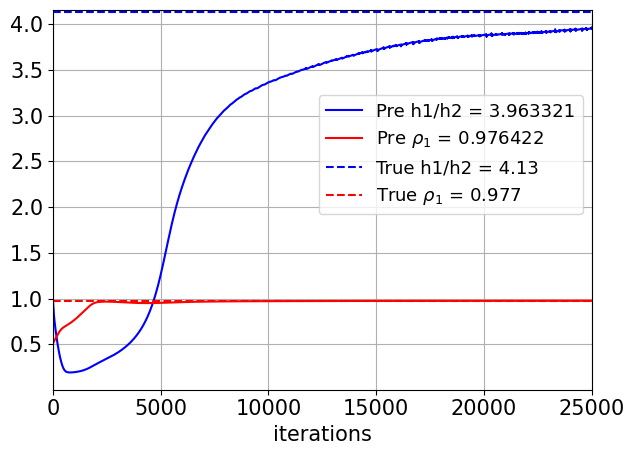}
        }
    \end{minipage}
    \caption{Effect of the number of spatial measurement points ($N_x$) on parameter estimation (Case (b): estimating $h_1/h_2$ and $\rho_1$). (a) Relative error for $h_1/h_2$ and $\rho_1$ as a function of $N_x$. (b)-(e) Convergence history for different $N_x$ values.}
    \label{Fig.3}
\end{figure*}

\begin{figure*}[t]
\centering
    \subfloat[Slice at $x=0$\label{fig4:a}]{
        \includegraphics[width=0.4\textwidth]{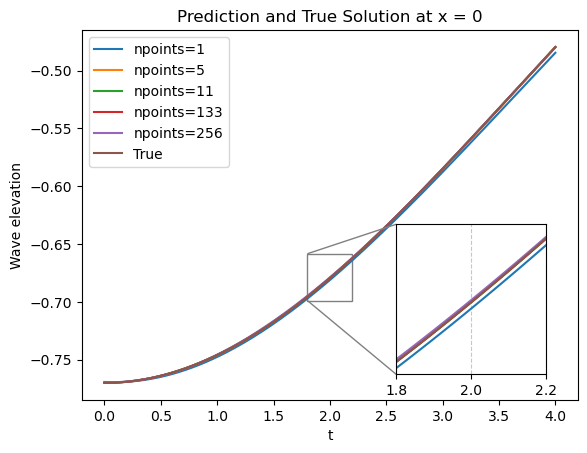}
    }
    \subfloat[Slice at $t=0$\label{fig4:b}]{
        \includegraphics[width=0.4\textwidth]{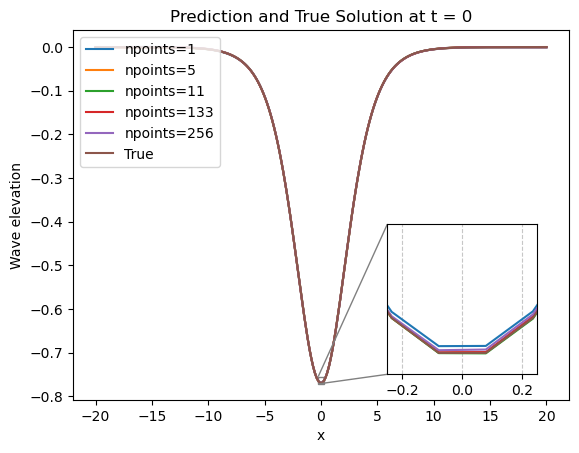}
    }
    \caption{Comparisons of predicted solutions (lines) versus exact solution (dots) along slices at (a) $x = 0$ (time evolution) and (b) $t = 0$ (spatial profile) for different numbers of spatial measurement points ($N_x$). Results correspond to parameter estimation Case (b).}
    \label{Fig.4}
\end{figure*}

Practical field measurements often suffer from sparse spatial sampling. To assess PINN’s robustness under data scarcity, we progressively reduce the number of measurement points $N_x$ along the x-axis from 256 (dense grid) down to 1 (single sensor), while maintaining temporal resolution ($N_t=201$ points in $t\in [0,4]$). The spatial points for $N_x < 256$ are chosen to be uniformly distributed within $x \in [-20, 20]$. The total number of data points is $N_d = N_x \times N_t$. We focus on Case (b) from Sec.~\ref{sec:param_select} (estimating $h_1/h_2$ and $\rho_1$). As shown in Fig.~\ref{Fig.3}, parameter estimation errors remain below $3\%$ even with only five symmetrically distributed sensors ($N_x=5$, corresponding to $x = -20, -10, 0, 10, 20$). Remarkably, single-point monitoring at $x=0$ ($N_x=1$) sustains $< 5\%$ error for $h_1/h_2$ and $< 8\%$ for $\rho_1$, showcasing the potential of PINNs under severe data limitation. outperforming conventional inversion methods under comparable conditions.
Cross-sectional comparisons at $x=0$ (time series) and $t=0$ (spatial profile) demonstrate excellent phase alignment and amplitude matching across different data sparsity levels in Fig.~\ref{Fig.4}, confirming that PINNs effectively leverage physics constraints to reconstruct the field from limited observations.

\subsubsection{Optimal Sensor Placement}\label{sec:placement}
\begin{figure*}[t]
    \centering
    \subfloat[Relative error vs. sensor location $x$\label{fig5:a}]{
        \includegraphics[width=0.48\textwidth]{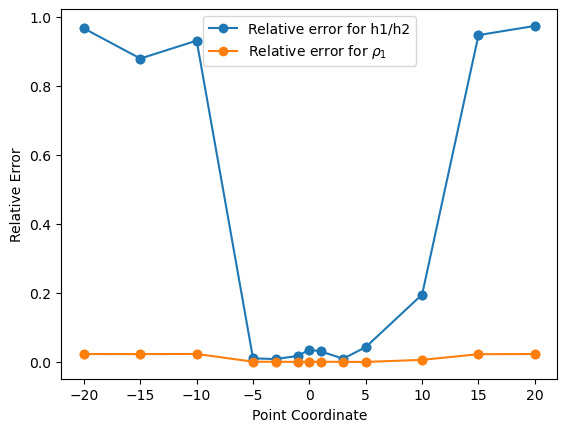}
    }
    \hfill
    \begin{minipage}[b]{0.48\textwidth}
        \centering
        \subfloat[$x=0$\label{fig5:b}]{
            \includegraphics[width=0.48\linewidth]{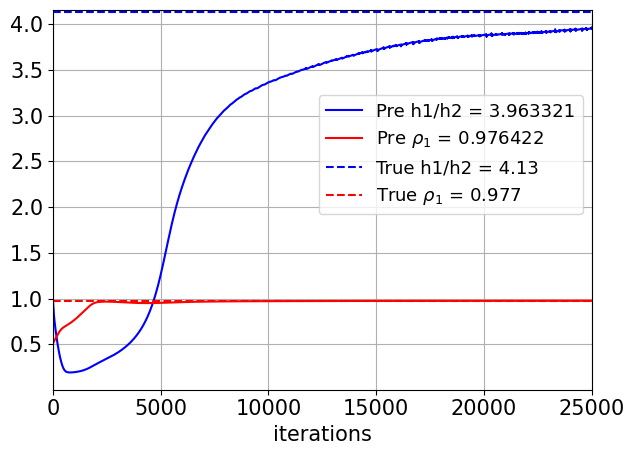}
        }
        \hfill
        \subfloat[$x=5$\label{fig5:c}]{
            \includegraphics[width=0.48\linewidth]{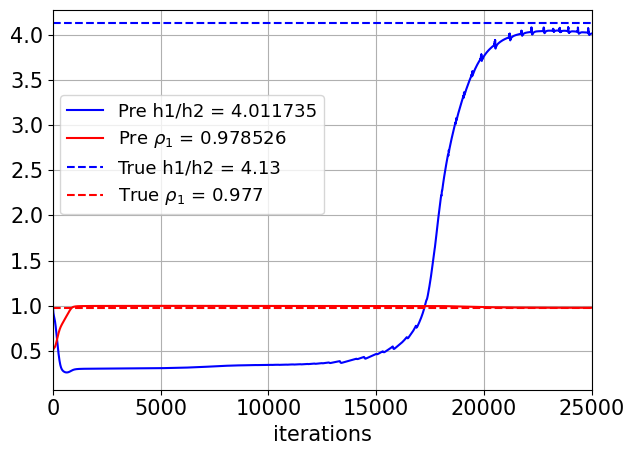}
        }
        \vfill
        \subfloat[$x=10$\label{fig5:d}]{
            \includegraphics[width=0.48\linewidth]{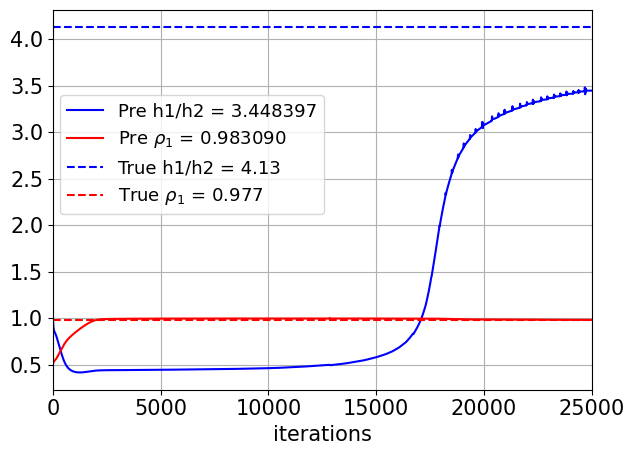}
        }
        \hfill
        \subfloat[$x=15$\label{fig5:e}]{
            \includegraphics[width=0.48\linewidth]{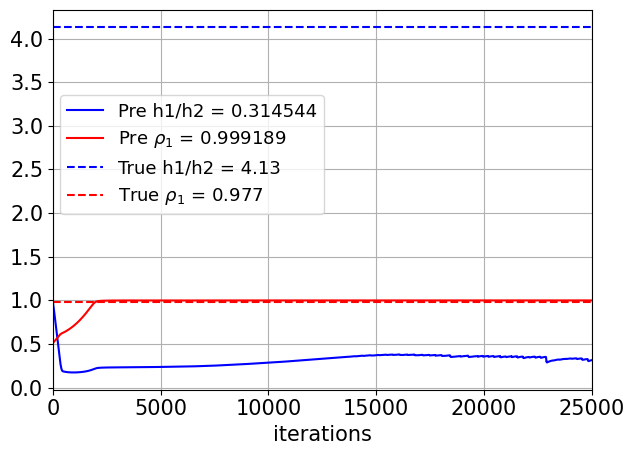}
        }
    \end{minipage}
    \caption{Impact of single sensor location ($x$) on parameter estimation (Case (b)). (a) Relative error for $h_1/h_2$ and $\rho_1$ as a function of the sensor's x-coordinate. (b)-(e) Convergence history for different sensor locations.}
    \label{Fig.5}
\end{figure*}

For single-sensor deployments ($N_x=1$), the spatial position ($x_d$) critically affects inversion accuracy. We systematically evaluate estimation errors by placing the single sensor at different locations $x \in[-20,20]$ for Case (b). As shown in Fig.~\ref{Fig.5}, the results identify an optimal measurement window roughly within $x \in [-5, 5]$. This region corresponds to where the solitary wave exhibits significant amplitude and temporal variation, thus providing the most information content for parameter inference. Errors escalate significantly for $|x| > 10$, where the wave amplitude is much smaller, making the measurements less sensitive to the underlying parameters and potentially more susceptible to noise (although noise was not included in this specific test). This suggests that strategic sensor placement should prioritize locations where the wave dynamics are most pronounced, specifically central waveform features instead of peripheral regions.

\subsubsection{Noise Robustness Analysis}\label{sec:noise_impact}
Previous sections investigated the influence of sensor location and sparsity on inversion results using noise-free data. In practical applications, measurement data inevitably contains noise. Therefore, evaluating the impact of noise on Physics-Informed Neural Network (PINN) inversion performance is of significant importance. Here, we specifically examine the effect of different noise levels on the relative errors of inversion results at a single observation point ($x=0$), corresponding to the optimal placement region identified in Sec.~\ref{sec:placement} and the $N_x=1$ case in Sec.~\ref{sec:sparsity}. We again focus on estimating $h_1/h_2$ and $\rho_1$ (Case (b)).

This study employs Gaussian noise added to the exact synthetic data $\eta(x_d^i, t_d^i)$ to simulate sensor noise: $\eta_{obs}(x_d^i, t_d^i) = \eta(x_d^i, t_d^i) + \epsilon_i$, where $\epsilon_i \sim \mathcal{N}(0, \sigma^2)$. This choice is based on two main considerations: (1) Gaussian noise is a common assumption for random errors, often justified by the Central Limit Theorem; (2) Measurement errors from many physical sensors approximate Gaussian distributions. In this work, the noise level is quantified by the standard deviation $\sigma$, which is set relative to the standard deviation of the clean data, $\sigma_{data} = \text{std}(\eta(x_d^i, t_d^i))$. Specifically, we define the noise standard deviation as $\sigma = \sigma_{data} \times \eta\%$, where $\eta\%$ denotes the noise level percentage parameter.

\begin{figure}
    \centering
    \includegraphics[width=0.8\linewidth]{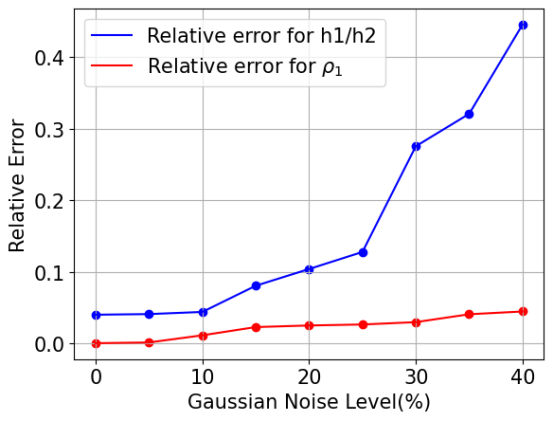}
    \caption{Impact of Gaussian noise level ($\eta\%$) on the relative error of inverted parameters ($h_1/h_2$ and $\rho_1$) using single-point observation data at $x=0$. Noise standard deviation $\sigma = \sigma_{data} \times \eta\%$.}
    \label{Fig.6}
\end{figure}

As illustrated in Fig.~\ref{Fig.6} (x-axis: Gaussian noise level $\eta\%$, y-axis: relative error), the inversion of $h_1/h_2$ and $\rho_1$ parameters exhibits distinct noise sensitivities. For $h_1/h_2$ inversion (blue curve), the relative error remains below 0.1 (10\%) when the noise level is under 20\%, demonstrating PINN's robustness for this parameter under moderate noise. Beyond a 20\% noise level, however, the relative error increases substantially. In contrast, $\rho_1$ inversion (red curve) maintains superior noise resistance, preserving low relative error (<0.05 or 5\%) even up to a 40\% noise level. These results confirm that the proposed PINN approach remains effective for parameter inversion under moderate noise conditions, although the sensitivity to noise can vary between different physical parameters.
 
\subsubsection{Discussion and Implications}\label{sec:discussion}
Our comprehensive analysis establishes PINNs as a potent tool for KdV-based parameter estimation, particularly in data-limited scenarios. Key insights include:
\begin{itemize}
\item[$\bullet$]\textbf{Parameter Coupling}: 
The KdV equation's mathematical structure , specifically the interdependence of physical parameters within the coefficients $c_0, c_1, c_2$, imposes inherent constraints on simultaneous multi-parameter estimation. Prior knowledge or fixing certain parameters (like $\rho_2$) significantly improves identifiability and necessitates prior knowledge of at least one parameter for stable inversion.
\item[$\bullet$]\textbf{Data Efficiency}: Physics regularization enables accurate parameter recovery with up to 99\% fewer measurements than traditional mesh-based methods require.
\item[$\bullet$]\textbf{Spatial Sensitivity}: 
 Measurement locations significantly impact inversion accuracy. Central measurement locations maximize information content due to stronger nonlinear interactions in core waveform regions, highlighting the importance of sensor placement strategy.
\item[$\bullet$]\textbf{Noise Robustness}: 
The PINN inversion exhibits reasonable robustness to moderate levels of Gaussian noise (e.g., up to 20-40\% relative noise standard deviation for single-point measurements at $x=0$), although different parameters show varying sensitivity. The lower layer density $\rho_1$ was found to be more robustly estimated than the layer height ratio $h_1/h_2$ in the presence of noise. This underscores the applicability of PINNs in realistic scenarios but also highlights the need to consider parameter-specific noise sensitivity.
\end{itemize}
These findings advance the application of deep learning in experimental fluid dynamics, particularly for stratified oceanographic studies where direct density profiling remains challenging. Future work should continue to investigate robustness with different noise types and potentially incorporate noise models explicitly, address multi-parameter estimation limitations through enhanced network architectures, more sophisticated loss weighting schemes, or hybrid data assimilation.

\section{Conclusion}\label{Conclusion}
This study has demonstrated the significant potential of Physics-Informed Neural Networks (PINNs) in solving both forward and inverse problems for the KdV equation for solving internal solitary waves in a two-layer fluid system, while developing efficient solution operators for parametric studies. Our comprehensive investigation yields several key insights that advance the field of scientific machine learning for nonlinear wave dynamics.

Through systematic numerical experiments, we have established that PINNs offer a robust framework for operator learning in KdV systems. The proposed architecture achieves remarkable accuracy (errors as low as $10^{-4}$) in predicting waveforms when sufficient training data is available. Our analysis reveals a clear logarithmic relationship between data quantity and solution accuracy, providing valuable guidance for experimental design in practical applications.

For inverse problem solving, PINNs exhibit exceptional performance in parameter estimation even with limited observational data. The method successfully recovers key physical parameters ($h_1/h_2$, $\rho_1$) with high accuracy from sparse, noise-free synthetic data, and demonstrates reasonable robustness when moderate levels of Gaussian noise are introduced. maintains estimation errors below 3\% with just five strategically placed sensors, and demonstrates particular effectiveness when measurements are concentrated in regions of strong nonlinear interaction (wave core region). The study highlights the importance of parameter identifiability, showing that fixing certain parameters significantly aids the recovery of others due to inherent coupling in the KdV formulation. Furthermore, we demonstrated the critical role of sensor placement, with measurements in regions of strong wave activity yielding superior results. These findings have immediate implications for field measurements in oceanographic research where density profile data is typically scarce and noisy. These findings significantly advance computational methods for nonlinear wave dynamics, offering practical solutions to long-standing challenges in ocean engineering and fluid dynamics where traditional methods face limitations in either efficiency or accuracy.

Future research could generalize this approach to handle more complex nonlinear models, including variable-coefficient KdV equations, coupled multi-physics problems involving wave-current-structure interactions, and higher-dimensional nonlinear wave phenomena. Investigating robustness to different types and levels of noise, potentially incorporating Bayesian PINNs for uncertainty quantification, and exploring adaptive sampling strategies remain crucial next steps. Furthermore, the application of PINNs to optimal control problems and real-time forecasting tools may open new possibilities for managing and predicting nonlinear wave phenomena.

\begin{acknowledgments}
R. Li and Z. Wang are supported by the National Natural Science Foundation of China with Grant Numbers 12402288 and 12202114, respectively. 
We thank Science42 Technology for their support.
\end{acknowledgments}

\section*{Data Availability Statement}
The data that support the findings of this study are available from the corresponding authors upon reasonable request.

\bibliography{references}

\providecommand{\noopsort}[1]{}\providecommand{\singleletter}[1]{#1}%
\begin{thebibliography}{27}%
\makeatletter
\providecommand \@ifxundefined [1]{%
 \@ifx{#1\undefined}
}%
\providecommand \@ifnum [1]{%
 \ifnum #1\expandafter \@firstoftwo
 \else \expandafter \@secondoftwo
 \fi
}%
\providecommand \@ifx [1]{%
 \ifx #1\expandafter \@firstoftwo
 \else \expandafter \@secondoftwo
 \fi
}%
\providecommand \natexlab [1]{#1}%
\providecommand \enquote  [1]{``#1''}%
\providecommand \bibnamefont  [1]{#1}%
\providecommand \bibfnamefont [1]{#1}%
\providecommand \citenamefont [1]{#1}%
\providecommand \href@noop [0]{\@secondoftwo}%
\providecommand \href [0]{\begingroup \@sanitize@url \@href}%
\providecommand \@href[1]{\@@startlink{#1}\@@href}%
\providecommand \@@href[1]{\endgroup#1\@@endlink}%
\providecommand \@sanitize@url [0]{\catcode `\\12\catcode `\$12\catcode `\&12\catcode `\#12\catcode `\^12\catcode `\_12\catcode `\%12\relax}%
\providecommand \@@startlink[1]{}%
\providecommand \@@endlink[0]{}%
\providecommand \url  [0]{\begingroup\@sanitize@url \@url }%
\providecommand \@url [1]{\endgroup\@href {#1}{\urlprefix }}%
\providecommand \urlprefix  [0]{URL }%
\providecommand \Eprint [0]{\href }%
\providecommand \doibase [0]{http://dx.doi.org/}%
\providecommand \selectlanguage [0]{\@gobble}%
\providecommand \bibinfo  [0]{\@secondoftwo}%
\providecommand \bibfield  [0]{\@secondoftwo}%
\providecommand \translation [1]{[#1]}%
\providecommand \BibitemOpen [0]{}%
\providecommand \bibitemStop [0]{}%
\providecommand \bibitemNoStop [0]{.\EOS\space}%
\providecommand \EOS [0]{\spacefactor3000\relax}%
\providecommand \BibitemShut  [1]{\csname bibitem#1\endcsname}%
\let\auto@bib@innerbib\@empty
\bibitem [{\citenamefont {Zhang}\ \emph {et~al.}(2023)\citenamefont {Zhang}, \citenamefont {Hu}, \citenamefont {Guo}, \citenamefont {Liang}, \citenamefont {Zhang}, \citenamefont {Chen}, \citenamefont {Xie},\ and\ \citenamefont {Du}}]{Zhang2023}%
  \BibitemOpen
  \bibfield  {author} {\bibinfo {author} {\bibfnamefont {M.}~\bibnamefont {Zhang}}, \bibinfo {author} {\bibfnamefont {H.~B.}\ \bibnamefont {Hu}}, \bibinfo {author} {\bibfnamefont {B.~B.}\ \bibnamefont {Guo}}, \bibinfo {author} {\bibfnamefont {Q.~Y.}\ \bibnamefont {Liang}}, \bibinfo {author} {\bibfnamefont {F.}~\bibnamefont {Zhang}}, \bibinfo {author} {\bibfnamefont {X.~P.}\ \bibnamefont {Chen}}, \bibinfo {author} {\bibfnamefont {Z.~L.}\ \bibnamefont {Xie}}, \ and\ \bibinfo {author} {\bibfnamefont {P.}~\bibnamefont {Du}},\ }\bibfield  {title} {\enquote {\bibinfo {title} {Predicting shear stress distribution on structural surfaces under internal solitary wave loading: {A} deep learning perspective},}\ }\href@noop {} {\bibfield  {journal} {\bibinfo  {journal} {Physics of Fluids}\ }\textbf {\bibinfo {volume} {36(3)}},\ \bibinfo {pages} {035153} (\bibinfo {year} {2023})}\BibitemShut {NoStop}%
\bibitem [{\citenamefont {Cheng}\ \emph {et~al.}(2024)\citenamefont {Cheng}, \citenamefont {Du}, \citenamefont {Wang}, \citenamefont {Xie}, \citenamefont {Hu}, \citenamefont {Chen}, \citenamefont {Li},\ and\ \citenamefont {Yuan}}]{Cheng2024}%
  \BibitemOpen
  \bibfield  {author} {\bibinfo {author} {\bibfnamefont {L.}~\bibnamefont {Cheng}}, \bibinfo {author} {\bibfnamefont {P.}~\bibnamefont {Du}}, \bibinfo {author} {\bibfnamefont {C.}~\bibnamefont {Wang}}, \bibinfo {author} {\bibfnamefont {Z.~L.}\ \bibnamefont {Xie}}, \bibinfo {author} {\bibfnamefont {H.~B.}\ \bibnamefont {Hu}}, \bibinfo {author} {\bibfnamefont {X.~P.}\ \bibnamefont {Chen}}, \bibinfo {author} {\bibfnamefont {Z.~Y.}\ \bibnamefont {Li}}, \ and\ \bibinfo {author} {\bibfnamefont {Z.~M.}\ \bibnamefont {Yuan}},\ }\bibfield  {title} {\enquote {\bibinfo {title} {Tuning control parameters of underwater vehicle to minimize the influence of internal solitary waves},}\ }\href@noop {} {\bibfield  {journal} {\bibinfo  {journal} {Ocean Engineering}\ }\textbf {\bibinfo {volume} {310}},\ \bibinfo {pages} {118681} (\bibinfo {year} {2024})}\BibitemShut {NoStop}%
\bibitem [{\citenamefont {Wang}, \citenamefont {Zhi},\ and\ \citenamefont {You}(2024)}]{Wang2024}%
  \BibitemOpen
  \bibfield  {author} {\bibinfo {author} {\bibfnamefont {R.~Z.}\ \bibnamefont {Wang}}, \bibinfo {author} {\bibfnamefont {C.~H.}\ \bibnamefont {Zhi}}, \ and\ \bibinfo {author} {\bibfnamefont {Y.~X.}\ \bibnamefont {You}},\ }\bibfield  {title} {\enquote {\bibinfo {title} {Numerical study of internal solitary wave loads on a submerged slender body with multi-parameter coupling},}\ }\href@noop {} {\bibfield  {journal} {\bibinfo  {journal} {Physics of Fluids}\ }\textbf {\bibinfo {volume} {36(11)}},\ \bibinfo {pages} {113361} (\bibinfo {year} {2024})}\BibitemShut {NoStop}%
\bibitem [{\citenamefont {Li}\ \emph {et~al.}(2023)\citenamefont {Li}, \citenamefont {Huang}, \citenamefont {Shi}, \citenamefont {Yang},\ and\ \citenamefont {Zhao}}]{Li2023}%
  \BibitemOpen
  \bibfield  {author} {\bibinfo {author} {\bibfnamefont {J.~M.~H.}\ \bibnamefont {Li}}, \bibinfo {author} {\bibfnamefont {X.~D.}\ \bibnamefont {Huang}}, \bibinfo {author} {\bibfnamefont {Y.}~\bibnamefont {Shi}}, \bibinfo {author} {\bibfnamefont {Y.~X.}\ \bibnamefont {Yang}}, \ and\ \bibinfo {author} {\bibfnamefont {W.}~\bibnamefont {Zhao}},\ }\bibfield  {title} {\enquote {\bibinfo {title} {Effects of internal solitary waves on three-dimensional sound propagation and {DOA} estimation in the {South} {China} {Sea}},}\ }\href@noop {} {\bibfield  {journal} {\bibinfo  {journal} {Applied Acoustic}\ }\textbf {\bibinfo {volume} {212}},\ \bibinfo {pages} {109612} (\bibinfo {year} {2023})}\BibitemShut {NoStop}%
\bibitem [{\citenamefont {Zheng}, \citenamefont {Lin},\ and\ \citenamefont {Chen}(2024)}]{Zheng2024}%
  \BibitemOpen
  \bibfield  {author} {\bibinfo {author} {\bibfnamefont {Y.~D.}\ \bibnamefont {Zheng}}, \bibinfo {author} {\bibfnamefont {J.}~\bibnamefont {Lin}}, \ and\ \bibinfo {author} {\bibfnamefont {X.}~\bibnamefont {Chen}},\ }\bibfield  {title} {\enquote {\bibinfo {title} {Application of fluctuations in the sound field in inversion of internal solitary wave phase speed},}\ }\href@noop {} {\bibfield  {journal} {\bibinfo  {journal} {Ocean Engineering}\ }\textbf {\bibinfo {volume} {305}},\ \bibinfo {pages} {117867} (\bibinfo {year} {2024})}\BibitemShut {NoStop}%
\bibitem [{\citenamefont {Brandt}\ and\ \citenamefont {Shipley}(2014)}]{Brandt2014}%
  \BibitemOpen
  \bibfield  {author} {\bibinfo {author} {\bibfnamefont {A.}~\bibnamefont {Brandt}}\ and\ \bibinfo {author} {\bibfnamefont {K.~R.}\ \bibnamefont {Shipley}},\ }\bibfield  {title} {\enquote {\bibinfo {title} {Laboratory experiments on mass transport by large amplitude mode-2 internal solitary waves},}\ }\href@noop {} {\bibfield  {journal} {\bibinfo  {journal} {Physics of Fluids}\ }\textbf {\bibinfo {volume} {26(4)}},\ \bibinfo {pages} {046601} (\bibinfo {year} {2014})}\BibitemShut {NoStop}%
\bibitem [{\citenamefont {Deepwell}\ and\ \citenamefont {Stastna}(2016)}]{Deepwell2016}%
  \BibitemOpen
  \bibfield  {author} {\bibinfo {author} {\bibfnamefont {D.}~\bibnamefont {Deepwell}}\ and\ \bibinfo {author} {\bibfnamefont {M.}~\bibnamefont {Stastna}},\ }\bibfield  {title} {\enquote {\bibinfo {title} {Mass transport by mode-2 internal solitary-like waves},}\ }\href@noop {} {\bibfield  {journal} {\bibinfo  {journal} {Physics of Fluids}\ }\textbf {\bibinfo {volume} {28(5)}},\ \bibinfo {pages} {056606} (\bibinfo {year} {2016})}\BibitemShut {NoStop}%
\bibitem [{\citenamefont {Helfrich}\ and\ \citenamefont {Melville}(2006)}]{Helfrich2006}%
  \BibitemOpen
  \bibfield  {author} {\bibinfo {author} {\bibfnamefont {K.~R.}\ \bibnamefont {Helfrich}}\ and\ \bibinfo {author} {\bibfnamefont {W.~K.}\ \bibnamefont {Melville}},\ }\bibfield  {title} {\enquote {\bibinfo {title} {Long nonlinear internal waves},}\ }\href@noop {} {\bibfield  {journal} {\bibinfo  {journal} {Annual Review of Fluid Mechanics}\ }\textbf {\bibinfo {volume} {38}},\ \bibinfo {pages} {395--425} (\bibinfo {year} {2006})}\BibitemShut {NoStop}%
\bibitem [{\citenamefont {Choi}(2022)}]{Choi2022}%
  \BibitemOpen
  \bibfield  {author} {\bibinfo {author} {\bibfnamefont {W.}~\bibnamefont {Choi}},\ }\bibfield  {title} {\enquote {\bibinfo {title} {High-order strongly nonlinear long wave approximation and solitary wave solution. {P}art 2. internal waves},}\ }\href@noop {} {\bibfield  {journal} {\bibinfo  {journal} {Journal of Fluid Mechanics}\ }\textbf {\bibinfo {volume} {952}},\ \bibinfo {pages} {A41} (\bibinfo {year} {2022})}\BibitemShut {NoStop}%
\bibitem [{\citenamefont {Benjamin}(1966)}]{Benjamin1966}%
  \BibitemOpen
  \bibfield  {author} {\bibinfo {author} {\bibfnamefont {T.~B.}\ \bibnamefont {Benjamin}},\ }\bibfield  {title} {\enquote {\bibinfo {title} {Internal waves of finite amplitude and permanent form},}\ }\href@noop {} {\bibfield  {journal} {\bibinfo  {journal} {Journal of Fluid Mechanics}\ }\textbf {\bibinfo {volume} {25}},\ \bibinfo {pages} {97--116} (\bibinfo {year} {1966})}\BibitemShut {NoStop}%
\bibitem [{\citenamefont {Grue}\ \emph {et~al.}(1999)\citenamefont {Grue}, \citenamefont {Jensen}, \citenamefont {Rusas},\ and\ \citenamefont {Sveen}}]{Grue1999}%
  \BibitemOpen
  \bibfield  {author} {\bibinfo {author} {\bibfnamefont {J.}~\bibnamefont {Grue}}, \bibinfo {author} {\bibfnamefont {A.}~\bibnamefont {Jensen}}, \bibinfo {author} {\bibfnamefont {P.}~\bibnamefont {Rusas}}, \ and\ \bibinfo {author} {\bibfnamefont {J.~K.}\ \bibnamefont {Sveen}},\ }\bibfield  {title} {\enquote {\bibinfo {title} {Properties of large-amplitude internal waves},}\ }\href@noop {} {\bibfield  {journal} {\bibinfo  {journal} {Journal of Fluid Mechanics}\ }\textbf {\bibinfo {volume} {380}},\ \bibinfo {pages} {257--278} (\bibinfo {year} {1999})}\BibitemShut {NoStop}%
\bibitem [{\citenamefont {Kodaira}\ \emph {et~al.}(2016)\citenamefont {Kodaira}, \citenamefont {Waseda}, \citenamefont {Miyata},\ and\ \citenamefont {Choi}}]{Kodaira2016}%
  \BibitemOpen
  \bibfield  {author} {\bibinfo {author} {\bibfnamefont {T.}~\bibnamefont {Kodaira}}, \bibinfo {author} {\bibfnamefont {T.}~\bibnamefont {Waseda}}, \bibinfo {author} {\bibfnamefont {M.}~\bibnamefont {Miyata}}, \ and\ \bibinfo {author} {\bibfnamefont {W.}~\bibnamefont {Choi}},\ }\bibfield  {title} {\enquote {\bibinfo {title} {Internal solitary waves in a two-fluid system with a free surface},}\ }\href@noop {} {\bibfield  {journal} {\bibinfo  {journal} {Journal of Fluid Mechanics}\ }\textbf {\bibinfo {volume} {804}},\ \bibinfo {pages} {201--223} (\bibinfo {year} {2016})}\BibitemShut {NoStop}%
\bibitem [{\citenamefont {Xuan}\ \emph {et~al.}(2024)\citenamefont {Xuan}, \citenamefont {Du}, \citenamefont {Wang}, \citenamefont {Peng},\ and\ \citenamefont {Wei}}]{Xuan2024}%
  \BibitemOpen
  \bibfield  {author} {\bibinfo {author} {\bibfnamefont {P.}~\bibnamefont {Xuan}}, \bibinfo {author} {\bibfnamefont {H.}~\bibnamefont {Du}}, \bibinfo {author} {\bibfnamefont {S.~D.}\ \bibnamefont {Wang}}, \bibinfo {author} {\bibfnamefont {P.}~\bibnamefont {Peng}}, \ and\ \bibinfo {author} {\bibfnamefont {G.}~\bibnamefont {Wei}},\ }\bibfield  {title} {\enquote {\bibinfo {title} {The applicability of nonlinear theories of the internal solitary wave and its loads on slender body by experimental methods},}\ }\href@noop {} {\bibfield  {journal} {\bibinfo  {journal} {Applied Ocean Research}\ }\textbf {\bibinfo {volume} {153}},\ \bibinfo {pages} {104279} (\bibinfo {year} {2024})}\BibitemShut {NoStop}%
\bibitem [{\citenamefont {Cuomo}\ \emph {et~al.}(2022)\citenamefont {Cuomo}, \citenamefont {Di~Cola}, \citenamefont {Giampaolo}, \citenamefont {Rozza}, \citenamefont {Raissi},\ and\ \citenamefont {Piccialli}}]{cuomoScientificMachineLearning2022}%
  \BibitemOpen
  \bibfield  {author} {\bibinfo {author} {\bibfnamefont {S.}~\bibnamefont {Cuomo}}, \bibinfo {author} {\bibfnamefont {V.~S.}\ \bibnamefont {Di~Cola}}, \bibinfo {author} {\bibfnamefont {F.}~\bibnamefont {Giampaolo}}, \bibinfo {author} {\bibfnamefont {G.}~\bibnamefont {Rozza}}, \bibinfo {author} {\bibfnamefont {M.}~\bibnamefont {Raissi}}, \ and\ \bibinfo {author} {\bibfnamefont {F.}~\bibnamefont {Piccialli}},\ }\bibfield  {title} {\enquote {\bibinfo {title} {Scientific {{Machine Learning Through Physics}}--{{Informed Neural Networks}}: {{Where}} we are and {{What}}'s {{Next}}},}\ }\href {\doibase 10.1007/s10915-022-01939-z} {\bibfield  {journal} {\bibinfo  {journal} {Journal of Scientific Computing}\ }\textbf {\bibinfo {volume} {92}},\ \bibinfo {pages} {88} (\bibinfo {year} {2022})}\BibitemShut {NoStop}%
\bibitem [{\citenamefont {Raissi}, \citenamefont {Perdikaris},\ and\ \citenamefont {Karniadakis}(2017{\natexlab{a}})}]{raissiPhysicsInformedDeep2017}%
  \BibitemOpen
  \bibfield  {author} {\bibinfo {author} {\bibfnamefont {M.}~\bibnamefont {Raissi}}, \bibinfo {author} {\bibfnamefont {P.}~\bibnamefont {Perdikaris}}, \ and\ \bibinfo {author} {\bibfnamefont {G.~E.}\ \bibnamefont {Karniadakis}},\ }\href {\doibase 10.48550/arXiv.1711.10561} {\enquote {\bibinfo {title} {Physics {{Informed Deep Learning}} ({{Part I}}): {{Data-driven Solutions}} of {{Nonlinear Partial Differential Equations}}},}\ } (\bibinfo {year} {2017}{\natexlab{a}}),\ \Eprint {http://arxiv.org/abs/1711.10561} {arXiv:1711.10561 [cs, math, stat]} \BibitemShut {NoStop}%
\bibitem [{\citenamefont {Raissi}, \citenamefont {Perdikaris},\ and\ \citenamefont {Karniadakis}(2017{\natexlab{b}})}]{raissiPhysicsInformedDeep2017a}%
  \BibitemOpen
  \bibfield  {author} {\bibinfo {author} {\bibfnamefont {M.}~\bibnamefont {Raissi}}, \bibinfo {author} {\bibfnamefont {P.}~\bibnamefont {Perdikaris}}, \ and\ \bibinfo {author} {\bibfnamefont {G.~E.}\ \bibnamefont {Karniadakis}},\ }\href {\doibase 10.48550/arXiv.1711.10566} {\enquote {\bibinfo {title} {Physics {{Informed Deep Learning}} ({{Part II}}): {{Data-driven Discovery}} of {{Nonlinear Partial Differential Equations}}},}\ } (\bibinfo {year} {2017}{\natexlab{b}}),\ \Eprint {http://arxiv.org/abs/1711.10566} {arXiv:1711.10566 [cs, math, stat]} \BibitemShut {NoStop}%
\bibitem [{\citenamefont {Raissi}, \citenamefont {Perdikaris},\ and\ \citenamefont {Karniadakis}(2019)}]{raissiPhysicsinformedNeuralNetworks2019}%
  \BibitemOpen
  \bibfield  {author} {\bibinfo {author} {\bibfnamefont {M.}~\bibnamefont {Raissi}}, \bibinfo {author} {\bibfnamefont {P.}~\bibnamefont {Perdikaris}}, \ and\ \bibinfo {author} {\bibfnamefont {G.}~\bibnamefont {Karniadakis}},\ }\bibfield  {title} {\enquote {\bibinfo {title} {Physics-informed neural networks: A deep learning framework for solving forward and inverse problems involving nonlinear partial differential equations},}\ }\href {\doibase https://doi.org/10.1016/j.jcp.2018.10.045} {\bibfield  {journal} {\bibinfo  {journal} {Journal of Computational Physics}\ }\textbf {\bibinfo {volume} {378}},\ \bibinfo {pages} {686--707} (\bibinfo {year} {2019})}\BibitemShut {NoStop}%
\bibitem [{\citenamefont {Moseley}, \citenamefont {Markham},\ and\ \citenamefont {{Nissen-Meyer}}(2020)}]{moseleySolvingWaveEquation2020}%
  \BibitemOpen
  \bibfield  {author} {\bibinfo {author} {\bibfnamefont {B.}~\bibnamefont {Moseley}}, \bibinfo {author} {\bibfnamefont {A.}~\bibnamefont {Markham}}, \ and\ \bibinfo {author} {\bibfnamefont {T.}~\bibnamefont {{Nissen-Meyer}}},\ }\href {\doibase 10.48550/arXiv.2006.11894} {\enquote {\bibinfo {title} {Solving the wave equation with physics-informed deep learning},}\ } (\bibinfo {year} {2020}),\ \bibinfo {note} {comment: 13 pages, 9 figures},\ \Eprint {http://arxiv.org/abs/2006.11894} {arXiv:2006.11894 [physics]} \BibitemShut {NoStop}%
\bibitem [{\citenamefont {Cai}\ \emph {et~al.}(2021{\natexlab{a}})\citenamefont {Cai}, \citenamefont {Wang}, \citenamefont {Wang}, \citenamefont {Perdikaris},\ and\ \citenamefont {Karniadakis}}]{caiPhysicsInformedNeuralNetworks2021a}%
  \BibitemOpen
  \bibfield  {author} {\bibinfo {author} {\bibfnamefont {S.}~\bibnamefont {Cai}}, \bibinfo {author} {\bibfnamefont {Z.}~\bibnamefont {Wang}}, \bibinfo {author} {\bibfnamefont {S.}~\bibnamefont {Wang}}, \bibinfo {author} {\bibfnamefont {P.}~\bibnamefont {Perdikaris}}, \ and\ \bibinfo {author} {\bibfnamefont {G.~E.}\ \bibnamefont {Karniadakis}},\ }\bibfield  {title} {\enquote {\bibinfo {title} {Physics-informed neural networks for heat transfer problems},}\ }\href {\doibase 10.1115/1.4050542} {\bibfield  {journal} {\bibinfo  {journal} {Journal of Heat Transfer}\ }\textbf {\bibinfo {volume} {143}},\ \bibinfo {pages} {060801} (\bibinfo {year} {2021}{\natexlab{a}})}\BibitemShut {NoStop}%
\bibitem [{\citenamefont {Bafghi}\ and\ \citenamefont {Raissi}(2023)}]{bafghiPINNsTF2FastUserFriendly2023}%
  \BibitemOpen
  \bibfield  {author} {\bibinfo {author} {\bibfnamefont {R.~A.}\ \bibnamefont {Bafghi}}\ and\ \bibinfo {author} {\bibfnamefont {M.}~\bibnamefont {Raissi}},\ }\href {\doibase 10.48550/arXiv.2311.03626} {\enquote {\bibinfo {title} {{{PINNs-TF2}}: {{Fast}} and {{User-Friendly Physics-Informed Neural Networks}} in {{TensorFlow V2}}},}\ } (\bibinfo {year} {2023}),\ \bibinfo {note} {comment: Accepted at Machine Learning and the Physical Sciences Workshop, NeurIPS 2023},\ \Eprint {http://arxiv.org/abs/2311.03626} {arXiv:2311.03626 [cs]} \BibitemShut {NoStop}%
\bibitem [{\citenamefont {Kadeethum}, \citenamefont {Jørgensen},\ and\ \citenamefont {Nick}(2020)}]{kadeethumPhysicsInformedNeuralNetworks2020}%
  \BibitemOpen
  \bibfield  {author} {\bibinfo {author} {\bibfnamefont {T.}~\bibnamefont {Kadeethum}}, \bibinfo {author} {\bibfnamefont {T.~M.}\ \bibnamefont {Jørgensen}}, \ and\ \bibinfo {author} {\bibfnamefont {H.~M.}\ \bibnamefont {Nick}},\ }\href {https://arxiv.org/abs/2005.09638} {\enquote {\bibinfo {title} {Physics-informed neural networks for solving inverse problems of nonlinear biot's equations: Batch training},}\ } (\bibinfo {year} {2020}),\ \Eprint {http://arxiv.org/abs/2005.09638} {arXiv:2005.09638 [physics.comp-ph]} \BibitemShut {NoStop}%
\bibitem [{\citenamefont {Cai}\ \emph {et~al.}(2021{\natexlab{b}})\citenamefont {Cai}, \citenamefont {Mao}, \citenamefont {Wang}, \citenamefont {Yin},\ and\ \citenamefont {Karniadakis}}]{caiPhysicsinformedNeuralNetworks2021}%
  \BibitemOpen
  \bibfield  {author} {\bibinfo {author} {\bibfnamefont {S.}~\bibnamefont {Cai}}, \bibinfo {author} {\bibfnamefont {Z.}~\bibnamefont {Mao}}, \bibinfo {author} {\bibfnamefont {Z.}~\bibnamefont {Wang}}, \bibinfo {author} {\bibfnamefont {M.}~\bibnamefont {Yin}}, \ and\ \bibinfo {author} {\bibfnamefont {G.~E.}\ \bibnamefont {Karniadakis}},\ }\bibfield  {title} {\enquote {\bibinfo {title} {Physics-informed neural networks ({{PINNs}}) for fluid mechanics: A review},}\ }\href {\doibase 10.1007/s10409-021-01148-1} {\bibfield  {journal} {\bibinfo  {journal} {Acta Mechanica Sinica}\ }\textbf {\bibinfo {volume} {37}},\ \bibinfo {pages} {1727--1738} (\bibinfo {year} {2021}{\natexlab{b}})}\BibitemShut {NoStop}%
\bibitem [{\citenamefont {Jin}, \citenamefont {Mattheakis},\ and\ \citenamefont {Protopapas}(2022)}]{jinPhysicsInformedNeuralNetworks2022}%
  \BibitemOpen
  \bibfield  {author} {\bibinfo {author} {\bibfnamefont {H.}~\bibnamefont {Jin}}, \bibinfo {author} {\bibfnamefont {M.}~\bibnamefont {Mattheakis}}, \ and\ \bibinfo {author} {\bibfnamefont {P.}~\bibnamefont {Protopapas}},\ }\bibfield  {title} {\enquote {\bibinfo {title} {Physics-informed neural networks for quantum eigenvalue problems},}\ }in\ \href {\doibase 10.1109/IJCNN55064.2022.9891944} {\emph {\bibinfo {booktitle} {2022 International Joint Conference on Neural Networks (IJCNN)}}}\ (\bibinfo {year} {2022})\ pp.\ \bibinfo {pages} {1--8}\BibitemShut {NoStop}%
\bibitem [{\citenamefont {Zhang}\ \emph {et~al.}(2022)\citenamefont {Zhang}, \citenamefont {Dao}, \citenamefont {Karniadakis},\ and\ \citenamefont {Suresh}}]{zhangAnalysesInternalStructures2022}%
  \BibitemOpen
  \bibfield  {author} {\bibinfo {author} {\bibfnamefont {E.}~\bibnamefont {Zhang}}, \bibinfo {author} {\bibfnamefont {M.}~\bibnamefont {Dao}}, \bibinfo {author} {\bibfnamefont {G.~E.}\ \bibnamefont {Karniadakis}}, \ and\ \bibinfo {author} {\bibfnamefont {S.}~\bibnamefont {Suresh}},\ }\bibfield  {title} {\enquote {\bibinfo {title} {Analyses of internal structures and defects in materials using physics-informed neural networks},}\ }\href {\doibase 10.1126/sciadv.abk0644} {\bibfield  {journal} {\bibinfo  {journal} {Science Advances}\ }\textbf {\bibinfo {volume} {8}},\ \bibinfo {pages} {eabk0644} (\bibinfo {year} {2022})}\BibitemShut {NoStop}%
\bibitem [{\citenamefont {{Escapil-Inchausp{\'e}}}\ and\ \citenamefont {Ruz}(2023)}]{escapil-inchauspeHAnalysisDataparallelPhysicsinformed2023}%
  \BibitemOpen
  \bibfield  {author} {\bibinfo {author} {\bibfnamefont {P.}~\bibnamefont {{Escapil-Inchausp{\'e}}}}\ and\ \bibinfo {author} {\bibfnamefont {G.~A.}\ \bibnamefont {Ruz}},\ }\bibfield  {title} {\enquote {\bibinfo {title} {H-{{Analysis}} and data-parallel physics-informed neural networks},}\ }\href {\doibase 10.1038/s41598-023-44541-5} {\bibfield  {journal} {\bibinfo  {journal} {Scientific Reports}\ }\textbf {\bibinfo {volume} {13}},\ \bibinfo {pages} {17562} (\bibinfo {year} {2023})}\BibitemShut {NoStop}%
\bibitem [{\citenamefont {Lu}\ \emph {et~al.}(2021)\citenamefont {Lu}, \citenamefont {Jin}, \citenamefont {Pang}, \citenamefont {Zhang},\ and\ \citenamefont {Karniadakis}}]{Lu_2021}%
  \BibitemOpen
  \bibfield  {author} {\bibinfo {author} {\bibfnamefont {L.}~\bibnamefont {Lu}}, \bibinfo {author} {\bibfnamefont {P.}~\bibnamefont {Jin}}, \bibinfo {author} {\bibfnamefont {G.}~\bibnamefont {Pang}}, \bibinfo {author} {\bibfnamefont {Z.}~\bibnamefont {Zhang}}, \ and\ \bibinfo {author} {\bibfnamefont {G.~E.}\ \bibnamefont {Karniadakis}},\ }\bibfield  {title} {\enquote {\bibinfo {title} {Learning nonlinear operators via deeponet based on the universal approximation theorem of operators},}\ }\href {\doibase 10.1038/s42256-021-00302-5} {\bibfield  {journal} {\bibinfo  {journal} {Nature Machine Intelligence}\ }\textbf {\bibinfo {volume} {3}},\ \bibinfo {pages} {218–229} (\bibinfo {year} {2021})}\BibitemShut {NoStop}%
\bibitem [{\citenamefont {Li}\ \emph {et~al.}(2021)\citenamefont {Li}, \citenamefont {Kovachki}, \citenamefont {Azizzadenesheli}, \citenamefont {Liu}, \citenamefont {Bhattacharya}, \citenamefont {Stuart},\ and\ \citenamefont {Anandkumar}}]{li2021fourierneuraloperatorparametric}%
  \BibitemOpen
  \bibfield  {author} {\bibinfo {author} {\bibfnamefont {Z.}~\bibnamefont {Li}}, \bibinfo {author} {\bibfnamefont {N.}~\bibnamefont {Kovachki}}, \bibinfo {author} {\bibfnamefont {K.}~\bibnamefont {Azizzadenesheli}}, \bibinfo {author} {\bibfnamefont {B.}~\bibnamefont {Liu}}, \bibinfo {author} {\bibfnamefont {K.}~\bibnamefont {Bhattacharya}}, \bibinfo {author} {\bibfnamefont {A.}~\bibnamefont {Stuart}}, \ and\ \bibinfo {author} {\bibfnamefont {A.}~\bibnamefont {Anandkumar}},\ }\href {https://arxiv.org/abs/2010.08895} {\enquote {\bibinfo {title} {Fourier neural operator for parametric partial differential equations},}\ } (\bibinfo {year} {2021}),\ \Eprint {http://arxiv.org/abs/2010.08895} {arXiv:2010.08895 [cs.LG]} \BibitemShut {NoStop}%
\end{thebibliography}%

\end{document}